\documentclass[11pt,a4paper]{article}

\usepackage[margin=2.5cm]{geometry}
\usepackage[T1]{fontenc}
\usepackage[utf8]{inputenc}
\usepackage{graphicx}
\usepackage{amsmath,amssymb}
\usepackage{comment}
\usepackage{multirow}
\usepackage{makecell}
\usepackage{xcolor}
\usepackage{authblk}
\usepackage{hyperref}  
\usepackage{aas_macro}
\usepackage[round,authoryear]{natbib}

\begin{document}

\title{Count-based spectral component imaging (CCI)\\
of solar flares in X-rays}

\author[1]{Alessia Guidetti}
\author[2,3]{Muriel Zo\"{e} Stiefel}
\author[2]{Paolo Massa}
\author[1]{Anna Maria Massone}
\author[1,4]{Michele Piana}
\author[2]{S\"{a}m Krucker}

\affil[1]{MIDA, Dipartimento di Matematica, Università di Genova, via Dodecaneso 35, 16146 Genova, Italy}
\affil[2]{University of Applied Sciences and Arts Northwestern Switzerland, Bahnhofstrasse 6, 5210 Windisch, Switzerland}
\affil[3]{ETH Z\"{u}rich, R\"{a}mistrasse 101, 8092 Z\"{u}rich, Switzerland}
\affil[4]{Istituto Nazionale di Astrofisica, Osservatorio Astrofisico di Torino, via Osservatorio 20, 10025 Pino Torinese, Italy}
\date{}

\maketitle

\begin{center}
\texttt{alessia.guidetti@edu.unige.it}
\end{center}

\begin{abstract}
X-ray emission in solar flares is produced by both multi-thermal plasma and accelerated electrons.
Classical imaging approaches reconstruct X-ray intensity maps which contain contributions from multiple spectral components (e.g., hot, superhot, and non-thermal), and do not allow retrieving the morphology of the different components separately.
We introduce a novel imaging technique, called ``Count-based spectral Component Imaging (CCI)'', to jointly reconstruct spatially resolved emission measure maps of the thermal components, and the electron flux distribution of the non-thermal component from data provided by the Spectrometer/Telescope for Imaging X-rays (STIX) aboard Solar Orbiter.
We formulate a linear model linking the Differential Emission Measure (DEM), approximated by two thermal components, and the non-thermal electron flux to the observed counts. The resulting inverse problem is solved with the Richardson-Lucy algorithm.
We apply CCI to STIX observations of SOL2024-10-01T22 and compare it with the previously developed Spectral Component Imaging (SCI) method, as well as classical imaging approaches. The reconstructed thermal and non-thermal components show good agreement with those obtained using SCI.
This proof-of-concept study shows that CCI obtains results consistent with SCI but with fewer inputs. In contrast with SCI, CCI can also be applied to hard X-ray focusing optics imaging.
\end{abstract}

\section{Introduction}

The hard X-ray (HRX) emission observed during solar flares has two main components: thermal radiation, produced by heated plasma, and non-thermal radiation, generated by accelerated particles \citep[e.g.,][]{1988psf..book.....T}. Further, 
observations reveal two distinct thermal plasma populations \citep[e.g.,][]{caspi2010rhessi,Stiefel_2025joint}: a hot component with temperature between \(\sim\)5 and \(\sim\)25 MK, and a superhot component with temperature $\gtrsim$30 MK.
The X-ray photon spectra produced by the two thermal components and, possibly, the non-thermal one have a significant overlap in energy, so it is challenging to image individual components from X-ray measurements.

Indirect X-ray imagers, such as the Reuven Ramaty Solar Spectroscopic Imager \citep[RHESSI;][]{Lin_2002}, the Spectrometer/Telescope for Imaging X-rays \citep[STIX;][]{krucker2020spectrometer} on board Solar Orbiter \citep{2020A&A...642A...1M}, and the Hard X-ray Imager \citep[HXI;][]{zhang2019hard} on board the ASO-S mission \citep{Gan_2019}, provide indirect imaging information by modulating the X-ray emission with pairs of tungsten grids, resulting in a set of Fourier components \citep[or visibilities; e.g.,][]{piana2022hard,massa2023stix}.
Hence, reconstructing the image of the flaring sources from RHESSI, STIX, and HXI observations in a specific energy range means inverting a limited number of Fourier data.
In the case where multiple spectral components contribute to the data recorded in the considered energy range, standard imaging techniques are unable to reconstruct each component separately.

To address this limitation, previous works \cite{Caspi_2015,stiefel2025spectral} proposed combining imaging and spectral information by means of three steps: a spectral fit of spatially integrated data to determine the parameters of thermal and non-thermal components; the determination of the visibility values of each component by solving a linear system whose coefficients are given by the spectral fit results; the image reconstruction of each component, separately. The approach by \citet{Caspi_2015} relies on the selection of specific energy ranges to define the linear system, and is therefore prone to inaccurate results in the case where the selected energy ranges are not suited.
In contrast, the Spectral Component Imaging \citep[SCI;][]{stiefel2025spectral} proposed using a broader set of energy channels. 
This approach increases photon statistics, exploits the full informational content of the data, and leads to more stable reconstructions. 

In this letter, we present the ``Count-based spectral Component Imaging (CCI)'' method, which uses STIX counts to simultaneously reconstruct the emission measure maps describing the thermal plasma within a simplified differential emission measure (DEM) model, and the map of the non-thermal electron flux (if present).
Formulating the imaging problem directly from STIX counts allows using the Richardson-Lucy algorithm \citep[RL; e.g.,][]{Massa_2019}, which has the advantage of accounting for the Poisson nature of the data noise and inherently imposes a positivity constraint on the reconstructed image.
The proposed technique requires fewer input parameters compared to the approach by \citet{Caspi_2015} and \citet{stiefel2025spectral} as the intensity of the emission measure and total electron flux integrated over the entire field of view are inferred from the data.
Further, our method consists of only two steps, i.e., spectral fit and simultaneous image reconstruction, and does not require solving a linear system of visibilities. We show that the CCI technique represents an effective alternative to previous approaches and allows an independent validation of the results obtained by SCI.

\section{Method}
In this Section, we derive the mathematical model relating the emission measure maps of the hot and superhot plasma components, together with the total electron flux map of a possible non-thermal component, to the counts recorded by STIX. This formulation leads to an inverse problem, whose solution enables the joint reconstruction of the thermal and non-thermal components from the observed STIX data.

\subsection{Imaging model}

The spatial distribution of the X-ray photon emission at energy
$\epsilon$ produced by thermal bremsstrahlung is
\begin{equation}\label{eq:ph_flux_model}
\varphi_{\rm th}(\mathbf{x};\epsilon)
=
\int
f_{\rm vth}(\epsilon,T)\,
{\rm DEM}(\mathbf{x};T)\,dT,
\end{equation}
where $\mathbf{x}=(x,y)$ denotes the position on the solar disk (with the coordinates expressed in arcseconds), $f_{\rm vth}(\epsilon,T)$ is the thermal bremsstrahlung model, and ${\rm DEM}(\mathbf{x};T)$ represents 
the differential emission measure at position $\mathbf{x}$ and temperature $T$.
In the following, we consider a simplified $\rm DEM$ model in which the plasma temperature takes only two values, $T_{\rm H}$ and $T_{\rm S}$, corresponding to the hot and super-hot components, respectively.
Under this assumption, we have
\begin{equation}\label{eq:dem_model}
{\rm DEM}(\mathbf{x}; T) = {\rm EM}_{\rm H}(\mathbf{x}) \, \delta_{T_{\rm H}}(T) + {\rm EM}_{\rm S}(\mathbf{x}) \, \delta_{T_{\rm S}}(T) ~,
\end{equation}
where ${\rm EM}_{\rm H}$ and ${\rm EM}_{\rm S}$ denote the emission measure maps at temperatures $T_{\rm H}$ and $T_{\rm S}$.
Hence, by plugging \eqref{eq:dem_model} into \eqref{eq:ph_flux_model}, the emitted photon flux reduces to the weighted sum 
\begin{equation}\label{eq:ph_th}
\varphi_{\rm th}(\mathbf{x}; \epsilon) = {\rm EM}_{\rm H}(\mathbf{x}) \,f_{\text{vth}}(\epsilon, T_{\rm H}) + {\rm EM}_{\rm S}(\mathbf{x}) \, f_{\text{vth}}(\epsilon, T_{\rm S})~.
\end{equation}
When non-thermal emission is also present, the photon flux includes a thick-target bremsstrahlung contribution.
The resulting photon flux is then given by
\begin{equation}\label{eq:ph_total}
\begin{aligned}
\varphi(\mathbf{x};\epsilon)=\varphi_{\rm th}(\mathbf{x}; \epsilon) + F(\mathbf{x})\, f_{\rm thick}(\epsilon;E_c,\delta)~,
\end{aligned}
\end{equation}
where $F$ denotes the total electron flux map, and $f_{\text{thick}}$ is the thick target bremsstrahlung model \citep[e.g.,][]{Brown_1971} that depends on the low-energy cutoff $E_c$ and spectral index $\delta$. The parameters $T_{\rm H}$, $T_{\rm S}$, $E_c$, and $\delta$ are inferred from a spectral fit analysis and must be provided
as input to our method. In the purely thermal case, the last term in \eqref{eq:ph_total} is omitted. 
Moreover, the proposed framework also naturally applies when only a single thermal component is considered.

The number of counts recorded by the $i$-th pixel in each STIX detector is
\begin{equation}\label{eq:fwd_model_continuous}
c_i(q) = \iint g_i(\mathbf{x})\, \varphi(\mathbf{x};\epsilon)\, {\rm SRM}(\epsilon,q) \,d\mathbf{x}\,d\epsilon  + b_i(q) ~,
\end{equation}
where $g_i(\mathbf{x})$ is the modulation pattern created by the grid pair in front of the pixel, the SRM is the Spectral Response Matrix describing the probability that a photon at energy $\epsilon$ is recorded as a count at energy $q$ and $b_i(q)$ is the intensity of the background emission detected at the same energy.

For the thermal components, define
\begin{equation}\label{eq:thermal_k}
k_{\rm th}(q,T) = \int f_{\rm vth}(\epsilon,T)\, {\rm SRM }(\epsilon,q)\,d\epsilon~,
\end{equation}
whereas for the non-thermal component define
\begin{equation}\label{eq:nonthermal_k}
k_{\rm nt}(q;E_c,\delta) =\int f_{\rm thick}(\epsilon;E_c,\delta)\,
{\rm SRM}(\epsilon,q)\,d\epsilon~.
\end{equation}
By plugging \eqref{eq:ph_total} into
\eqref{eq:fwd_model_continuous} we obtain
\begin{equation}\label{eq:fwd_model_components}
\begin{aligned}
c_i(q) = & \int g_i(\mathbf{x}) \, {\rm EM}_{\rm H}(\mathbf{x}) \, k_{\rm th}(q,T_{\rm H}) \,d\mathbf{x} \\
&+ \int g_i(\mathbf{x}) \,
{\rm EM}_{\rm S}(\mathbf{x})\, k_{\rm th}(q,T_{\rm S}) \,d\mathbf{x} \\
&+ \int g_i(\mathbf{x}) \, F(\mathbf{x})\, k_{\rm nt}(q;E_c,\delta) \,d\mathbf{x} + b_i(q) ~ .
\end{aligned}
\end{equation}
Equation \eqref{eq:fwd_model_components} represents the linear forward model relating the emission measure maps (and, when present, the non-thermal electron flux map) to the counts recorded by STIX detector pixels at different energies.

\subsection{Discretization and inversion}

Given the STIX detector pixel measurements at different count energies, our goal is to retrieve the maps ${\rm EM}_{\rm H}$ and ${\rm EM}_{\rm S}$ of the hot and superhot thermal components, and the map $F$ of the total electron flux (if a non-thermal component is present). 
Once the spatial domain and the energy axis are discretized in \eqref{eq:fwd_model_components}, the reconstruction problem can be written as
\begin{equation}
C = G M K + B~.
\end{equation}
Here, $C$ is the matrix of counts measured by detector pixels in the
count-energy channels, i.e., $C_{ij}=c_i(q_j)$;
$B$ contains the background counts detected by the STIX pixels at all energies, with $B_{ij}=b_i(q_j)$; $G$ accounts for the spatial modulation of the STIX grids, with
$G_{ip}=g_i(\mathbf{x}_p)$; and $K$ is the discretization of either the thermal bremsstrahlung or the thick target bremsstrahlung model convolved with the SRM. 
The $j$-th column of $K$ is defined as $K_{j} = \left[k_{\rm th}(q_j , T_{\rm H}), k_{\rm th}(q_j ,T_{\rm S}), k_{\rm nt}(q_j ; E_c,\delta)
\right]^T$.
The matrix $M$, to be retrieved, consists of the emission measure maps and the total electron flux map, i.e., the $p$-th row of $M$ is given by
$M_{p} = [{\rm EM}_{\rm H}(\mathbf{x}_p) , {\rm EM}_{\rm S}(\mathbf{x}_p), F(\mathbf{x}_p)]$.

To reconstruct $M$ from $C$ we employ the well-known Richardson-Lucy (RL) algorithm, which is the same as the Expectation Maximization method presented in \citet{Massa_2019} for the solution of the standard imaging problem from STIX data. In this Letter, we prefer to address to this standard algorithm as RL in order to avoid confusion between Emission Measure and Expectation Maximization. 
The iterative RL step is given by 
\begin{equation}
\label{eq:ml_em_algorithm}
M^{(k+1)} =
M^{(k)} \odot
\frac{
   G^{T} \!\left(\dfrac{C}{G M^{(k)} K + b}\right) K^{T}
}{
   G^{T}\mathbf{1}\, K^{T}
}~,
\end{equation}

\noindent
where $\odot$ denotes element-wise multiplication and the division is also
element-wise. 
The RL algorithm is initialized with ${M^{(0)} = \mathbf{1}}$.
In the following, we refer to the proposed reconstruction approach as the CCI imaging technique since it jointly reconstructs the thermal components as their emission measure maps, and the non-thermal component as its total electron flux map.

\section{Application to STIX data}

\begin{figure}[!h]
\centering
\includegraphics[width=0.8\linewidth]{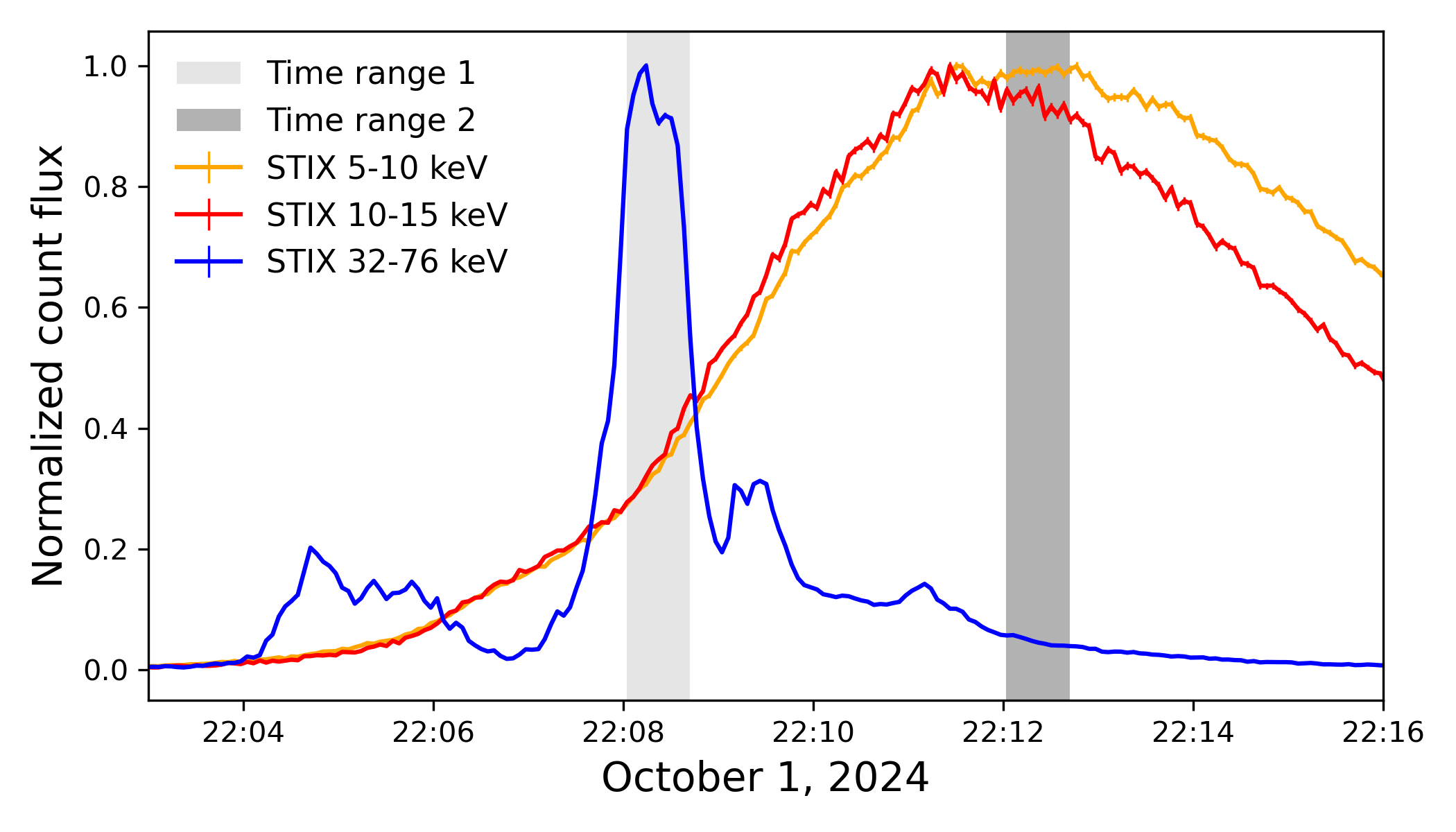}
\caption{Normalized lightcurves registered by STIX during the SOL2024-10-01T22.
The 5-10 keV and 10-15 keV lightcurves (orange and red lines, respectively) are measured by the Background detector \citep[BKG;][]{Stiefel_2025bkg}, while the 32-76 keV lightcurve (blue line) is provided by the imaging sub-collimators.
The time intervals considered for our analysis are highlighted by the gray shaded areas.}
\label{fig:lightcurve}
\end{figure}

In this Section we apply the CCI technique to STIX observations of the SOL2024-10-01T22 flare \citep[e.g.,][]{Matsumoto_2025,stiefel2025spectral,Matsumoto_2026}.
The GOES class of this event is X7.1.
We consider two different time ranges, 22:08:02-22:08:42~UT and 22:12:02-22:12:42 UT (where the times are expressed as arrival time at Solar Orbiter), which correspond to the peak of the non-thermal and of the thermal emission, respectively.
In the following, we refer to these two time intervals as time range 1 and time range 2.
The time evolution of the considered event, as well as the selected time ranges, are displayed in Fig.~\ref{fig:lightcurve}.

\begin{figure*}[t]
\centering
\includegraphics[width=\linewidth]{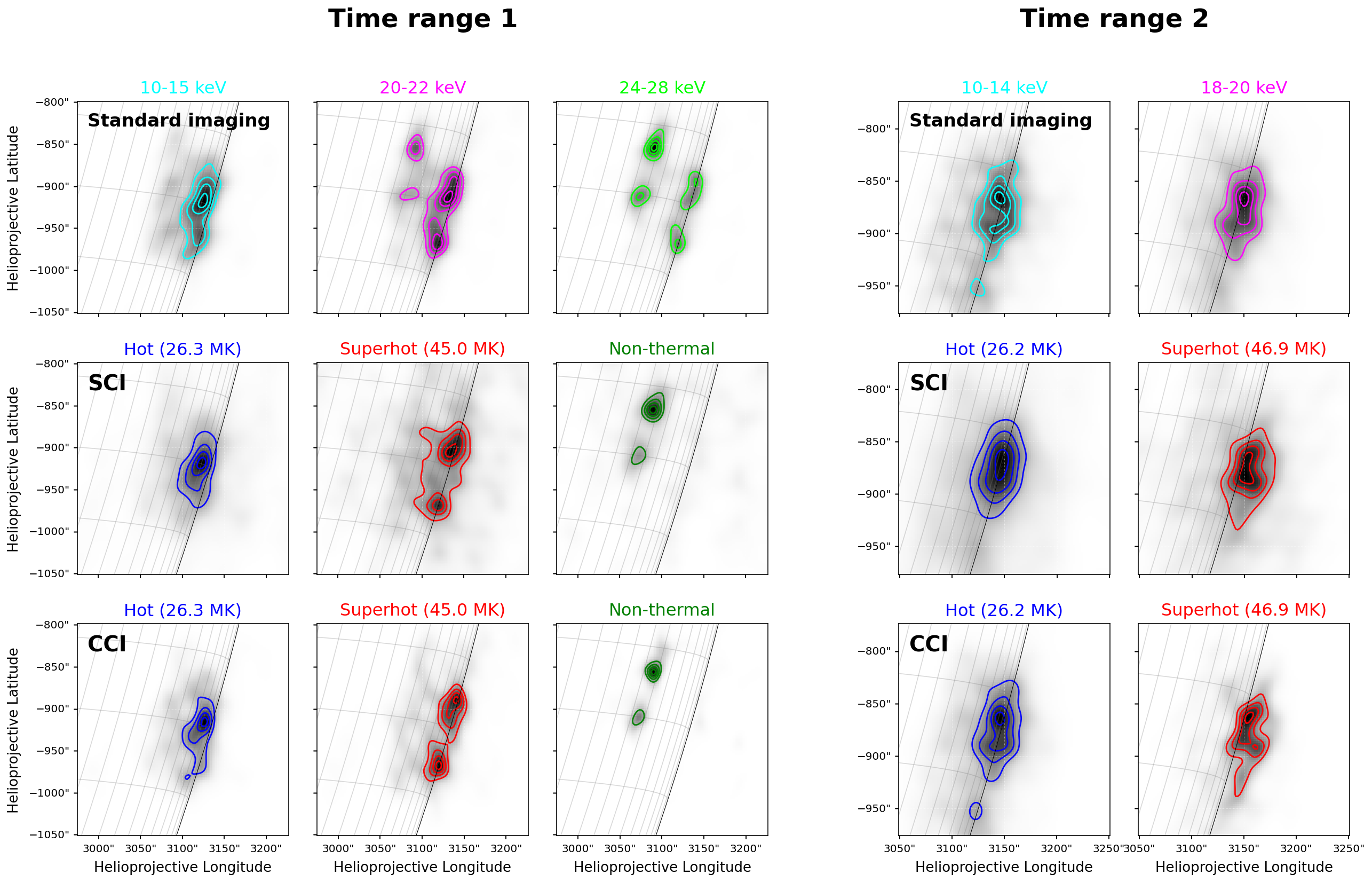}
\caption{Imaging results for the two analyzed time intervals.
The first row shows RL reconstructions obtained in different energy bands.
The second and third rows report the component-based reconstructions obtained with the SCI and CCI methods, respectively.
In each panel, contour levels at 30\%, 50\%, 70\%, and 90\% of the peak intensity value are plotted.
}
\label{fig:test_22_08}
\end{figure*}

\subsection{Spectral analysis}\label{spectral_an}

During the two considered time ranges, the attenuator is inserted for STIX. 
For the spectral analysis of the time steps we therefore used the approach of joint spectral fitting using the STIX background detector and the imaging detectors \citep{Stiefel_2025joint}. 
For examples of STIX spectral fits for this flare, we refer to \citet{stiefel2025spectral}. Spectral analysis shows that the first time range contains two thermal components with temperatures $T_H=26.3 \pm 0.1$ MK and $T_S=45.0 \pm 0.8$ MK, and a non-thermal component with spectral index $\delta = 3.16 \pm 0.01$  and low-energy cutoff $ E_c=26.8 \pm 0.7$ keV.
The second time range consists of two thermal components with $T_H = 26.2 \pm 0.1$ MK and $T_S=46.9 \pm 0.3$ MK.
The total EM and total electron flux values of each fitted component are reported in the third column of Table~\ref{tab:comparison_time_ranges}.
In analogy with SCI, an albedo component is fitted in the spectral analysis of both time ranges.
However, as for SCI and standard imaging from energy ranges, albedo is neglected in the image reconstruction process.

\begin{table}[h]
\centering
\setlength{\tabcolsep}{8pt}
\renewcommand{\arraystretch}{1.3}

\label{tab:comparison_time_ranges}

\resizebox{\columnwidth}{!}{%
\begin{tabular}{|c|l|c|c|c|c|}
\hline
\multirow{2}{*}{TR} & \multirow{2}{*}{Estimated parameter} & \multirow{2}{*}{\makecell{Spectral \\ analysis}} & \multicolumn{3}{c|}{CCI} \\
\cline{4-6}
 &  &  & 0\% & 5\% & 10\% \\
\hline

\multirow{3}{*}{1}
& Total EM hot  
& $3.04 \pm 0.03$ 
& $3.35 \pm 0.08$
& $2.86 \pm 0.06$
& $2.59 \pm 0.05$ \\

& Total EM superhot 
& $0.076 \pm 0.009$ 
& $0.106 \pm 0.010$
& $0.090 \pm 0.008$
& $0.077 \pm 0.007$ \\

& Tot. integrated el. flux  
& $2.03 \pm 0.12$ 
& $2.74 \pm 0.05$
& $1.84 \pm 0.11$
& $1.59 \pm 0.09$ \\

\hline

\multirow{2}{*}{2}
& Total EM hot 
& $8.37 \pm 0.07$ 
& $9.58 \pm 0.11$
& $8.26 \pm 0.10$
& $7.32 \pm 0.08$ \\

& Total EM superhot
& $0.088 \pm 0.004$ 
& $0.121 \pm 0.004$
& $0.097 \pm 0.003$
& $0.085 \pm 0.002$ \\

\hline
\end{tabular}}
\caption{Comparison between the parameter values provided by the spectral fit and those retrieved from the images obtained with CCI.  The considered time range (TR) is indicated in the leftmost column. The units of the total EM values are $10^{49}$ cm$^{-3}$, while the units of the total integrated electron flux are $10^{35}$ electrons s$^{-1}$. 
For CCI, the estimated quantities are computed using three different intensity thresholds, namely 0\%, 5\%, and 10\% of the peak intensity.}
\end{table}

\subsection{Proof of concept}
Figure~\ref{fig:test_22_08} shows the imaging results provided by RL (in selected energy bands), SCI, and CCI for the two time intervals.
Standard energy-band reconstructions generally contain contributions from multiple spectral components, as particularly evident in the 20-22 keV image of time interval 1,
where structures associated with both thermal and non-thermal emission are visible simultaneously, making the physical interpretation of the observed morphology ambiguous. 
In contrast, both CCI and SCI directly recover the individual thermal and non-thermal components.

The reconstructions obtained with the CCI approach are in good agreement with those derived using SCI. In particular,
the location and morphology of the hot, super-hot, and non-thermal components are consistent across the two methods.
However, our CCI technique does not require the total emission measures and the total electron flux to be provided as input. 
Rather, these quantities can be inferred by summing the pixel values of the resulting images.
This is in contrast with SCI, which utilizes these values to compute the relative contribution of each spectral component to the visibilities.
Table~\ref{tab:comparison_time_ranges} reports the total emission measures of the hot and super-hot components, together with the total electron flux, when present, as retrieved from the CCI reconstructions.
To compute these quantities from the CCI images, we summed the pixel values within three different level curves, which correspond to 0\%, 5\% and 10\% of the peak intensity.
This is done to assess how much including low intensity pixels, which likely contain reconstruction artifacts, affect the estimation of the quantities of interest.
For comparison, Table~\ref{tab:comparison_time_ranges} reports the parameter values obtained with spectral analysis.

The imaging model \eqref{eq:fwd_model_components} takes as inputs the temperature values of the hot and super-hot components, and the spectral index and low-energy cutoff of the non-thermal component, when present.
To estimate the uncertainty on the results reported in Table~\ref{tab:comparison_time_ranges}, we generated 100 values of input parameters according to the spectral fit results reported in Section~\ref{spectral_an}.
In particular, we perturbed the input parameters with gaussian noise where the mean is set equal to the fitted values and the standard deviation is set equal to the corresponding uncertainty.
Finally, we performed 100 reconstructions from the perturbed input values using \eqref{eq:ml_em_algorithm}, and we computed the standard deviation of the resulting total emission measure and total electron flux values.  
Table~\ref{tab:comparison_time_ranges} shows that the parameter values obtained using the 0\% and the 10\% intensity threshold, respectively, overestimate and underestimate those retrieved by spectral fits. 
In the worst case scenario, the relative differences are up to 40\%.
Considering the 5\% intensity threshold yields a better agreement, with relative differences within 10\%, except for the EM of the superhot component of time range 1, where the discrepancy is 18\%.
In fact, pixel values below 5\% of the peak intensity cannot be reliably retrieved given the limited dynamic range of STIX (typically 1:20).
However, considering the 10\% level appears to be a conservative choice.
Overall there is a good agreement between the CCI results and those from spectral analysis, given also the accuracy with which these parameters can be constrained from STIX observations.

\section{Conclusions}
In this work we introduced the CCI imaging tecnique for the joint reconstruction of spatially resolved emission measure maps of the hot and super-hot thermal components, together with the electron flux distribution of the non-thermal component. 
Compared to existing methods, the proposed approach relies on fewer assumptions: only the temperatures of the thermal components and, when present, low energy cut-off and spectral index of the non-thermal one, are required as input parameters. The total emission measures values and total electron flux are instead directly estimated through the reconstruction process. 
Moreover, once the input parameters are specified, the CCI approach provides the reconstructions in a single step, without requiring the intermediate estimation of visibilities followed by image reconstruction, as done for SCI. 
We showed that the reconstructions obtained with CCI are in good agreement with those provided by the SCI method, and that the estimated total EM and total integrated electron flux values are consistent with spectral fit results.
Overall, the CCI imaging technique provides a self-consistent and complementary framework for the analysis of STIX observations.
In contrast with SCI, the CCI technique can be applied for the analysis of hard X-ray focusing optics data \citep[e.g.][]{Glesener_2017}.
In this case, image deconvolution using the instrument point spread function should replace the image reconstruction task from modulated STIX counts.   

\section*{Acknowledgements}
The STIX instrument is an international collaboration between Switzerland, Poland, France, Czech Republic, Germany, Austria, Ireland, and Italy.
PM and SK are supported by Swiss PRODEX grant for STIX. 
MZS is supported by the Swiss National Science Foundation Grant 2000-1-240022.
AMM and AG acknowledge the support of the PRIN 2022 Project `Greedy Optimal Sampling for Solar Inverse Problems (GOSSIP)' 2022HFB32T, CUP: D53C24003380006. The research by AG, AMM and MP was performed within the framework of the MIUR Excellence Department Project awarded to Dipartimento di Matematica, Università di Genova, CUP: D33C23001110001.

\bibliographystyle{plainnat}
\bibliography{references}

\end{document}